\documentclass{aa}
\usepackage{natbib}
\usepackage{gensymb}
\usepackage{graphicx}
\usepackage{color}
\usepackage{txfonts}
\usepackage{ulem}
\usepackage{multirow}
\usepackage[breaklinks, colorlinks, citecolor=blue]{hyperref} 
\definecolor{orange}{RGB}{255,127,0}
\usepackage{lipsum}

\newcommand{\tab}[1]{Table\,\ref{#1}}
\newcommand{\fig}[1]{Fig.\,\ref{#1}}
\newcommand{\sect}[1]{Section\,\ref{#1}}
\newcommand{\eq}[1]{Eq.\,(\ref{#1})}
\newcommand{\eqn}[1]{\eq{#1}}

\newcommand{\ith}{$i$th}

\newcommand{\kth}{$k$th}

\begin{document}

   \title{Non-ideal magnetohydrodynamics of self-gravitating filaments}
    \titlerunning{Non-ideal magnetohydrodynamics of self-gravitating filaments}
    \authorrunning{Guti\'errez-Vera}
   \author{Nicol Guti\'errez-Vera\inst{1}
            \and
          Tommaso Grassi\inst{2}
          \and
          Stefano Bovino\inst{1}
          \and
          Alessandro Lupi\inst{3,4}
          \and
          Daniele Galli\inst{5}
          \and
          Dominik R.G. Schleicher\inst{1}
          }

   \institute{Departamento de Astronomía, Facultad Ciencias Físicas y Matemáticas, Universidad de Concepción, Av. Esteban Iturra s/n Barrio Universitario, Casilla 160, Concepción, Chile,
              \email{ngutierrezv@udec.cl}
             \and
              Centre for Astrochemical Studies, Max-Planck-Institut f\"ur Extraterrestrische Physik, Gie\ss nbachstr. 1, 85749 Garching bei M\"unchen, Germany
              \and
              Dipartimento di Fisica ``G. Occhialini'', Universit\'a degli Studi Milano-Bicocca, Piazza della Scienza 3, I-20126 Milano, Italy
              \and
              INFN, Sezione di Milano-Bicocca, Piazza della Scienza 3, I-20126 Milano, Italy
              \and
              INAF-Osservatorio Astrofisico di Arcetri, Largo E. Fermi 5, I-50125 Firenze, Italy
             }

   \date{Received ; accepted }

  \abstract
   {Filaments have been studied in detail through observations and simulations. A range of numerical works have separately investigated how chemistry and diffusion effects, as well as magnetic fields and their structure impact the gas dynamics of the filament. However, non-ideal effects have hardly been explored thus far.}
   {We investigate how non-ideal magnetohydrodynamic (MHD) effects, combined with a simplified chemical model affect the evolution and accretion of a star-forming filament.}
   {We modeled an accreting self-gravitating turbulent filament using \textsc{lemongrab}, a one-dimensional (1D) non-ideal MHD code that includes chemistry. We explore the influence of non-ideal MHD, the orientation and strength of the magnetic field, and the cosmic ray ionization rate, {on} the evolution of the filament, with particular focus on the width and accretion rate.}
   {We find that the filament width and the accretion rate are determined by the magnetic field properties, including the initial strength, the coupling with the gas controlled by the cosmic ray ionization rate, and the orientation of the magnetic field with respect to the accretion flow direction. Increasing the cosmic-{ray} ionization rate leads to a behavior closer to that of ideal MHD, reducing the magnetic pressure support and, hence, damping the accretion efficiency with a consequent broadening of the filament width. For the same reason, we obtained a narrower width and a larger accretion rate when we reduced the initial magnetic field strength. Overall, while these factors affect the final results by approximately a factor of~2, removing the non-ideal MHD effects results in a much greater variation (up to a factor of~7).}
   {The inclusion of non-ideal MHD effects and the cosmic-ray ionization is crucial for the study of self-gravitating filaments and in determining critical observable quantities, such as the filament width and accretion rate.}


   \maketitle

\section{Introduction}
Filamentary structures in molecular clouds (e.g., \citealt[][for the Polaris Flare cloud]{Ward2010}, \citealt[][for the Taurus region]{Kirk2013}, and \citealt[][for the Chamaeleon cloud complex]{Alves2014}) have been revealed by \textit{Herschel} observations \citep{Andre2010,Molinari2010}, suggesting that pre-stellar cores may form from the gravitational fragmentation of marginally supercritical and magnetized filaments \citep{Konyves2015, Benedettini2018}. The role of the magnetic field in the star formation process, and particularly of its orientation with respect to the filament axis, has been widely discussed \citep{Soler2013, PlanckCollab2016XXXIII} and it represents a crucial point in understanding how the cores embedded in the filament grow in mass and trigger the formation of protostellar objects. Observational data from Zeeman-effect surveys \citep{Crutcher2012} show that the maximum strength of the interstellar magnetic field is $\sim 10$~$\mu$G for gas with densities below $n_{\rm H} \sim 300$ cm$^{-3}$, increasing significantly at higher densities. This increase is determined by the balance between ambipolar diffusion and the accumulation of the magnetic field due to condensation. The comparison between magnetized and non-magnetized models, either in equilibrium \citep{Toci2015a, Toci2015b} or dynamical \citep{Hennebelle2008}, shows that magnetic fields have a strong influence on the filamentary structure and the fragmentation process within the filaments \citep{Tilley2007,Kirk2015}. A relevant driver of these processes is the geometry of the magnetic field.

Polarization measurements often show that the orientation of the magnetic field is nearly perpendicular to the major axis of the star-forming filament, but aligned with low-density filamentary structures  {\citep[e.g.,][]{Soler2013, Plankcollab2016, Jow2018, Soler2019}}, a result also found in numerical simulations {\citep[e.g.,][]{Nakamura2008, Soler2017, Seifried2020, Koertgen2020}}. \citet{Cox2016} analyzed the filamentary structure of the Musca cloud, differentiating the main filament and the low-density filamentary structures close to it, showing that they are parallel to the plane-of-the-sky local magnetic field and quasi-perpendicular to the main filament. Additional observations revealed a large-scale network of sub-filaments connected to the filamentary structures \citep[e.g.,][]{Schneider2010}, aligned with the direction of the magnetic field, which can represent a mass reservoir for further growth of the filament and the cores embedded into it. 

The accretion of filaments is important in the context of star formation, because it could play a key role in preventing dense filaments from collapsing to spindles, and in maintaining constant widths during the evolutionary process \citep{ANDRE2017}. Some authors have focused on different ways in which the ambient material could be accreted, while considering the constraints imposed by the magnetic field direction inside molecular clouds. Based on \textit{Herschel} observations, \citet{Shimajiri2019} showed that the accretion of ambient gas can be driven by the gravitational potential of the filament. This provides a strong support for the scenario of mass accretion along magnetic field lines (oriented nearly perpendicular to the filament) proposed by \citet{Palmeirim2013}. Assuming a magnetic field perpendicular to the filament, \citet{Hennebelle&Andre2013} considered radial accretion in a self-gravitating filament due to the combination of accretion-driven turbulence and dissipation of the turbulence by ion-neutral friction. \citet{Gomez2022} proposed that the width measured by observations may change depending on the tracer used and then suggested a tracer-dependent estimate of the accretion rate onto the filament. In \citet{Gomez2014}, a similar behavior was shown, whereby filaments were proposed to be river-like flow structures, with the gas falling onto the filament changing direction as the gas density increases, and accreting mainly in the perpendicular direction. 

Magnetic fields also have  an impact on the characteristic width of the filament. From an analysis of Herschel observations, \citet{Arzoumanian2011} identified a peaked distribution of filament widths around 0.1~pc, including both low-column density and star-forming filaments. Whether or not this width is a universal filament property is still a matter of debate \citep[e.g.,][]{Smith2014, Panopoulou2017, Hacar2022}, but several theoretical works have attempted to explain it. For example, in simulations of idealized filaments, \citet{Seifried2015} found that the 0.1~pc width \citep{Arzoumanian2011} assumed in their initial conditions could be maintained if the magnetic field is longitudinal.

Ideal magnetohydrodynamics (MHD) may not be sufficient for studying filaments, due to the importance of ambipolar diffusion and other non-ideal processes \citep{Grassi2019}. The coupling between magnetic field and gas depends on the density and the ionization degree \citep{Shu1983}. The resistivity coefficient of ambipolar diffusion is indeed strongly dependent on the abundance of ions (e.g., \citealt{Grassi2019}), with collisions between charged dust grains and neutral gas particles dominating the momentum transfer. In this context, cosmic rays play a fundamental role as they ionize molecular clouds \citep{Padovani2009, Padovani2011}. \citet{Chen2014} simulated core formation in colliding flows including ambipolar diffusion, finding a transient ambipolar diffusion phase during the formation of the shock layer that allows for the formation of cores with higher mass-to-flux ratio. In a subsequent work, \citet{Ntormousi2016} reported wider filaments as a consequence of non-ideal effects. However, numerical simulations including ambipolar diffusion in MHD turbulence have not determined that it plays a role in setting a characteristic scale \citep{Oishi2006, Burkhart2015}.

In this paper, we explore how different physical parameters, such as the magnetic field strength and cosmic ray ionization rate, affect the evolution and accretion of a self-gravitating filament, including chemistry and non-ideal effects. In Sect. 2, we introduce our initial conditions and the non-ideal magnetohydrodynamic equations, together with details on the chemistry and microphysics employed in this work. In Sect. 3, we discuss the reference model and the parameter study, along with the impact of those parameters on the accretion process. In Sect. 4, we present a discussion of the main results referring to the accretion rates and filament widths. Finally, in Sect. 5, we discuss some limitations of our approach and in Sect. 6, we summarize our main conclusions.

\section{Methodology}

In the following, we describe the initial conditions and the theoretical framework adopted in this work.

\subsection{Initial conditions}

Our model consists of a self-gravitating turbulent filament with 0.1~pc width in a 2~pc size periodic box sampled with 1024~cells, which can accrete mass from both sides of the $x$-axis, namely,~the coordinate on which the hydrodynamical equations are solved. For our study, we model the filament as a nonuniform slab with an oblique magnetic field.

\begin{figure}
    \centering
    \includegraphics[width=0.49\textwidth]{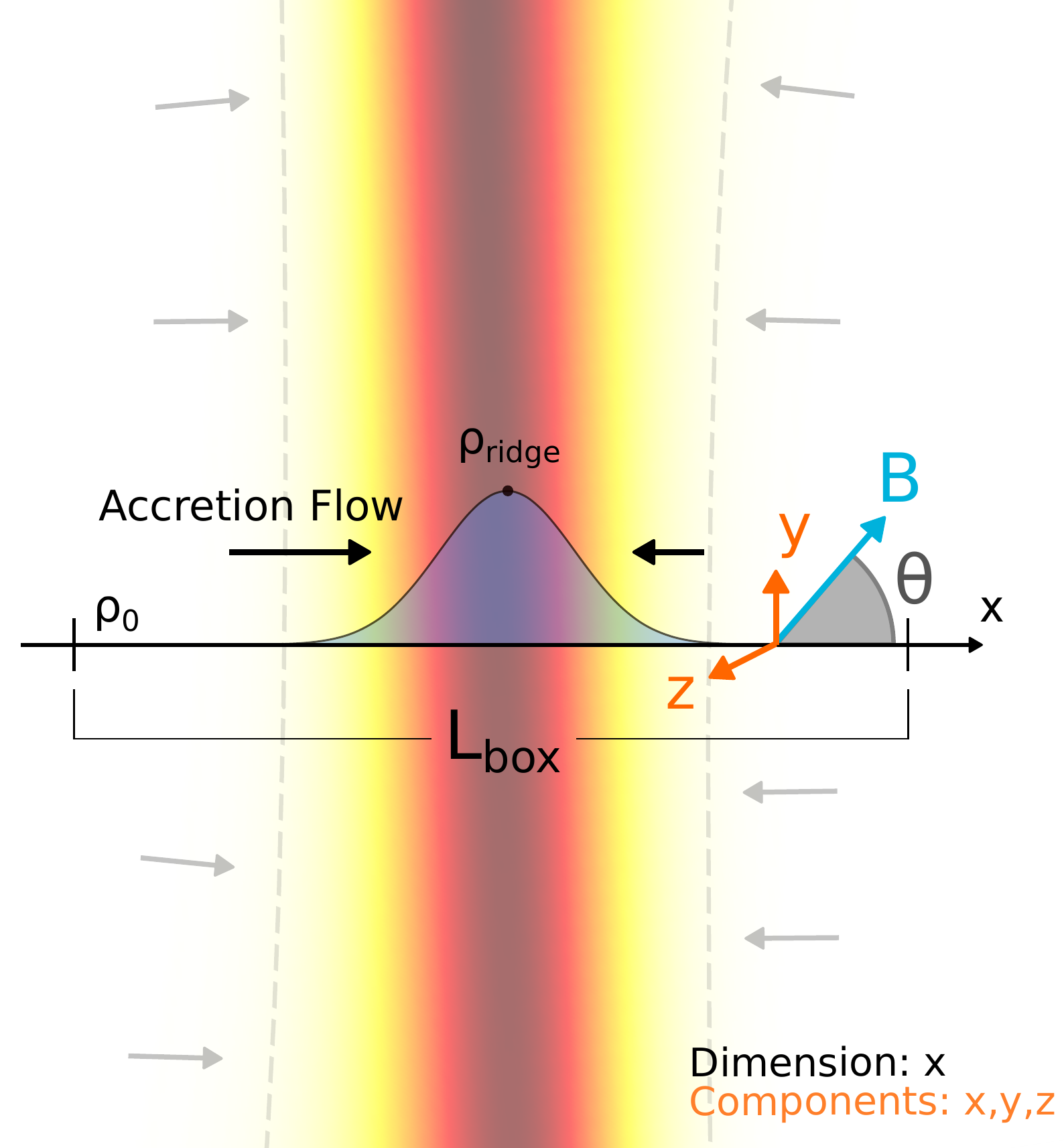}
    \caption{Representative sketch of our set-up. The code evolves a vector variable with three~components ($x,\,y,\,z,$) along one coordinate ($x$); hence, $y$ and $z$ have periodic boundary conditions. The $x$ coordinate traverses the filament (shaded gradient) and the initial density distribution along this component follows a Plummer-like profile with maximum density, $\rho_{\rm ridge}$, and background density, $\rho_0$, see \eqn{eq:plummer}. The gas accretes along the same axis, as indicated by the arrows. The magnetic field is inclined by an angle $\theta$ with respect to the $x$-axis, where  $\theta=\arctan{B_y/B_x}$. We note that the color palette employed for the filament is only included for illustration purposes.}
    \label{fig:model}
\end{figure}

To model the filament, we adopted a Plummer-like profile as shown in \citet{Arzoumanian2011}, given by:
\begin{equation}\label{eq:plummer}
 \rho(x) = \frac{\rho_\mathrm{ridge}}{\left[1+(x/x_{\rm flat})^2\right]^{p/2}},
\end{equation}
where $\rho_{\rm ridge}$ is the central density of the filament, $x_{\rm flat}$ is the characteristic width of the flat inner plateau of the filament, and $p$ is the typical exponent of the profile. For this study, we assumed the same parameters as adopted in \citet{Kortgen2018}, namely,~$p=2$ and $x_{\rm flat}= 0.033$~pc, and we aligned the minor axis of the filament with the $x$-axis. Figure~\ref{fig:model} shows a sketch of the filament setup.

The box is initialized with a uniform temperature of 15~K and is filled with a low-density medium with ${\rho_{\rm 0} \sim 2\times 10^{-21}}$~g~cm$^{-3}$ outside the filament region, with the latter modeled according to the Plummer-like profile in \eq{eq:plummer}, where we set $\rho_\mathrm{ridge} \sim 3\times10^{-19}$~g~cm$^{-3}$. The mass per unit length ratio is calculated as $(M/L) = 3\,(M/L)_{\rm crit}$ with $(M/L)_{\rm crit} = 2c_{\rm s}^2 / G$ \citep{Inutsuka1992} and $c_{\rm s} = \sqrt{k_{\rm B}T/\mu m_{\rm p}}$ is the speed of sound, with $G$, $k_{\rm B}$, $\mu$ and $m_{\rm p}$ being the gravitational constant, the Boltzmann constant, the mean molecular weight, and the mass of the proton, respectively. Then, $(M/L) \sim 86$~M$_\odot$~pc$^{-1}$. {We note that we initialize the filament at the critical line mass to guarantee an axial collapse.}

The free-fall time of the background gas, defined as:
\begin{equation}
    t_{\rm ff} = \sqrt{\frac{3\pi}{32G\rho_0}},
\end{equation}
corresponds to $\sim 1.3$~Myr.

The filament is initialized including turbulence with Mach number $\mathcal{M}=2$ {in the $\varv_x, \varv_y$, and $\varv_z$ components of the velocity field}, following a Burgers-like power spectrum, as described in \citet{Bovino2019} \citep[see also][]{Kortgen2017}. Given the randomness of the initial velocity field, to avoid any net momentum that could produce a drift  in the filament along the direction of the gravitational force, we subtract the mean initial velocity of the $x$-component.

\subsection{Magnetic field orientation}

The magnetic field ${\bf B}$ is set perpendicular to the $z$-axis and is tilted relative to the $x$-axis (the accretion flow direction) by an angle $\theta = \arctan(B_y/B_x)$, as shown in \fig{fig:model}. We note that when $\theta=0$ the field is parallel to the flow direction, while for $\theta=\pi/2$ the field is perpendicular to the flow. Given the dimensionality of the code {and the implicit assumption that} $\partial_y=\partial_z=0$, the solenoidal condition requires $\partial_xB_x=0$. Hence, $B_x$ remains constant throughout the box, whereas the initial condition of the $y$ component of the field is assumed to scale as \citep{Crutcher2010}:
\begin{equation}\label{crutcher}
 B_y = B_{y,0} \left(\frac{\rho}{\rho_0}\right)^k,
\end{equation}
where $k=0.5$, and $B_{y,0} \equiv B_0\sin{\theta}$, with $B_0=10~\mu$G. We note that since $B_x$ is constant by construction, $\theta$ also varies with $\rho$, indicating that $\theta$ in our initial conditions refers to $B_y = B_{y,0}$.

\subsection{Numerical framework}\label{sect:framework}

To study the evolution of the filament, we employ the \textsc{lemongrab} code \citep{Grassi2019}, a 1D code that solves the non-ideal MHD equations time-dependently. The code also includes cooling and heating processes as well as grain chemistry (including charged grains). For the purposes of this study, we added self-gravity to \textsc{lemongrab}, as described below.

The coupled non-ideal MHD equations can be written as:
\begin{eqnarray}
&& \partial_t \,\rho = -\partial_x [\rho \varv_x],\\
&& \partial_t [\rho \varv_x] = -\partial_x \left[\rho \varv_x^2 + P^* - \frac{B_x^2}{4\pi}\right] - \rho\partial_x \Phi, \\
&& \partial_t [\rho \varv_y] = -\partial_x \left[\rho \varv_x\varv_y - \frac{B_xB_y}{4\pi}\right], \\
&& \partial_t [\rho \varv_z] = -\partial_x \left[\rho \varv_x\varv_z - \frac{B_xB_z}{4\pi}\right],
\end{eqnarray}
\begin{eqnarray}
&& \partial_t B_x = 0,\\
&& \partial_t B_y = -\partial_x \left[\varv_xB_y - \varv_yB_x + \frac{\eta_{\rm AD}}{|{\bf B}|^2} (F_{B,x}B_y-F_{B,y}B_x) \right],\\
&& \partial_t B_z = -\partial_x \left[\varv_xB_z - \varv_zB_x + \frac{\eta_{\rm AD}}{|{\bf B}|^2} (F_{B,z}B_x-F_{B,x}B_z) \right],
\end{eqnarray}
\begin{eqnarray}
&& \partial_t {\cal E} = \Gamma_{\rm cr}-\Lambda-\rho \varv_x\partial_x \Phi-\partial_x \left\lbrace ({\cal E}+P^*)\varv_x - \frac{B_x}{4\pi}({\bf v}\cdot{\bf B}) \right. \nonumber\\
&& \left.-\frac{\eta_{\rm AD}}{\pi|{\bf B}|^2}[(F_{B,z}B_x-F_{B,x}B_z)B_z
- (F_{B,x}B_y-F_{B,y}B_x)B_y] \right\},
\end{eqnarray}
\begin{eqnarray}
&& \partial_t (\rho X_i) = -\partial_x (\rho X_i \varv_x) + \mathcal{P}_i - \rho X_i \mathcal{L}_i.
\end{eqnarray}
In the above equations, $\Phi$ is the gravitational potential, $\Lambda$ is the cooling rate, $\Gamma_{\rm cr}$ is the cosmic ray heating rate, and $\mathcal{P}_i$ and $\mathcal{L}_i$ are the production and loss terms of the $\ith$ chemical species.

The total pressure $P^\star$ is:
\begin{equation}
    P^\star = P + \frac{|{\bf B}|^2}{8\pi}.
\end{equation}
We assume an ideal equation of state for the thermal pressure:
\begin{equation}
    {\cal E} = \frac{P}{\gamma-1} + \frac{\rho|{\bf v }|^2}{2} + \frac{|{\bf B}|^2}{8\pi},
\end{equation}
and related to the temperature $T$, needed by the chemistry, via the ideal gas law
\begin{equation}
    P = \frac{\rho k_{\rm B}}{\mu m_{\rm p}}T.
\end{equation}
The components of the Lorentz force are
\begin{eqnarray}
    F_{B,x} &=& -B_y\cdot \partial_x B_y -  B_z\cdot \partial_x B_z,\\
    F_{B,y} &=& B_x\cdot \partial_x B_y,\\
    F_{B,z} &=& B_x\cdot \partial_x B_z.
\end{eqnarray}

\noindent The ambipolar diffusion resistivity coefficient is 
\begin{equation}
    \eta_{\rm AD} = c^2 \left( \frac{\sigma_{\rm P}}{\sigma_{\rm P}^2 + \sigma_{\rm H}^2} - \frac{1}{\sigma_{||}} \right),
\end{equation}
where $\sigma_{\rm P}$, $\sigma_{\rm H}$, and $\sigma_{||}$ are, respectively, the Pedersen, Hall, and parallel conductivities, and $c$ is the speed of light.

Given the spatial discretization of the code, the Poisson equation can be easily discretized over the 1D grid at second-order via central differencing, as
\begin{equation}
    \partial^2_x\Phi \approx \frac{\Phi_{k+1}-\Phi_{k}+\Phi_{k-1}}{\Delta x^2} = 4\pi{\rm G}\rho_k,
\end{equation}
where $\Phi_k$ and $\rho_k$ are the gravitational potential and density of the $\kth$ cell (ranging from 1 to $N$, with $N$ as the number of resolution elements in the simulation), and $\Delta x$ the spatial resolution of the simulation. For the boundary cells, we have included different boundary conditions (periodic, Dirichlet, or outflowing) among which the user can arbitrarily choose; although in this work we only consider periodic ones, that is, $\Phi_1 = \Phi_N$. The potential is then calculated via matrix inversion, and the gravitational acceleration is finally obtained via central differencing as:
\begin{equation}
g_x = -\partial_x \Phi \approx -\frac{\Phi_{k+1}-\Phi_{k-1}}{2\Delta x}.
\end{equation}

In this work, we employ a cooling function which depends on both temperature and number density. It corresponds to conditions of collisional ionization equilibrium typical of the interstellar gas,  determined using \textsc{cloudy} \citep{Ferland1998} and tabulated by \citet{Shen2013}. We imposed a temperature floor of 10~K by including an artificial heating term, defined as $\Lambda(T=10\,$K$)$, and added to the $\partial_t {\cal E}$ equation.

For details of the calculation of pressure, heating, cosmic ray ionization rate, conductivities, and the chemical evolution of the species, we refer to \citet{Grassi2019}. In this work, we employed their reduced chemical network, including eight~species: electrons e$^-$, X, X$^+$, and neutral and charged grains, namely:~g, g$^+$, g$^{++}$, g$^-$, g$^{--}$; The species X and X$^+$ are a proxy of all neutrals and cations produced by a chain of fast reactions following H$_2$ ionization \citep[see][for details]{Fujii2011,Grassi2019}. We assumed that the mass of X is the same as the mass of H$_2$, and the dust grains rates were weighted assuming a MRN distribution \citep{Mathis1977} with the same characteristics as in \citet{Grassi2019}. The cosmic ray ionization rate of X is assumed to be initially uniform at a value $\zeta_{\rm cr}=10^{-17}$~s$^{-1}$. We set the initial electron fraction to $f_i = n_{\rm e^-}/n_{\rm H_2} = 10^{-7}$ and the dust-to-gas mass ratio to ${\mathcal D} = 10^{-2}$, as listed in  \tab{tab:params}.

\section{Results}
In this section, we present the results of our simulations divided in four parts. In \sect{reference}, we present our reference model based on the initial conditions listed in \tab{tab:params}. In \sect{magnetic field}, we show the effects of varying the orientation and strength of the magnetic field. Separately, in \sect{CRIR}, we consider different cosmic rays ionization rate and, finally, in \sect{other parameters} we explore the effect of other physical quantities such as the initial turbulence seed, Mach number, density regimes, and $k$ exponent of the magnetic field-density relation, namely, Eq. \ref{crutcher}.%\eqn{crutcher}.
%\LEt{ Parentheses are not needed around the equation number.}
\subsection{Reference case} \label{reference}

\begin{center}
\begin{table}
    \begin{tabular}{lll}
        \hline \hline
        Physical quantity & Numerical value & Units\\
        \hline 
        $\rho_\mathrm{ridge}$ &$3.424\times10^{-19}$ & g cm$^{-3}$\\ 
        $x_{\rm flat}$ & 0.0333 & pc \\
        $L_{\text{box}}$ & 2 & pc \\
        $T$ & 15 & K\\
        $\mathcal{M}$ & 2 & -\\
        $B_0$ & 10 & $\mu$G \\
        $B_{x,0}$ & 2.25 & $\mu$G \\
        $B_{y,0}$ & 9.74 & $\mu$G \\
        $B_z$ & 0 & $\mu$G \\

        $f_i$ & $10^{-7}$ & -\\
        $\zeta_{\rm cr}$ & $10^{-17}$ & s$^{-1}$ \\
        $\mathcal{D}$ & $10^{-2}$ & - \\
        $\varv_x, \varv_y, \varv_z$ & see text & - \\
        \hline
    \end{tabular}
    \caption{Initial conditions for the filament setup of the reference case.}
    \label{tab:params}
\end{table}
\end{center}

\begin{figure*}
    \centering
    \includegraphics[scale=0.55]{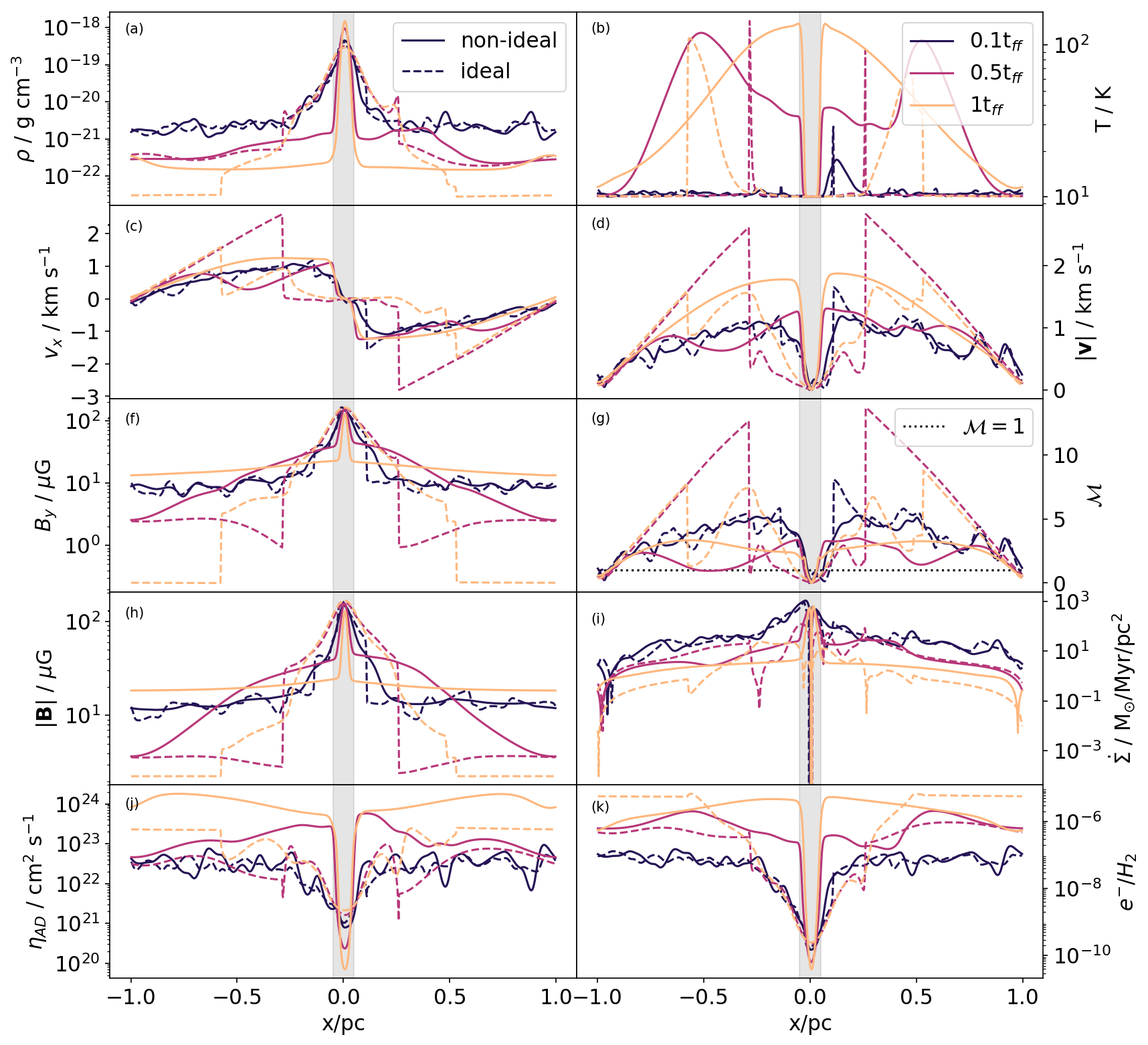}
    \caption{Profile along the $x$-axis of various physical quantities of the model at three different times: $t/t_{\rm ff}=$ 0.1, 0.5 and 1.0. ({\it a}\/) density $\rho$, ({\it b}\/) temperature $T$, ({\it c}\/) $x$-component of the velocity $\varv_x$, ({\it d}\/) total velocity {magnitude} |{v}|, ({\it e}\/) $y$-component of magnetic field $B_y$, ({\it f}\/) Mach number $\mathcal{M}$, ({\it g}\/) total magnetic field |{B}| {magnitude}, ({\it h}\/) mass flux $\dot{\Sigma}$, ({\it i}\/) ambipolar diffusion resistivity coefficient $\eta_{\rm AD}$,  and ({\it j}\/) abundance $e^-/$H$_2$. The solid lines correspond to non-ideal MHD, dashed lines to ideal MHD. The gray area in the middle corresponds to 0.1 pc to guide the reader to identify the typical width of the filaments. We note that in panel ({\it j}\/), in the ideal case (dashed lines), $\eta_{\rm AD}$ is reported for the sake of comparison, but it is not included in the MHD equations.}
    \label{fig:refPrincipal}
\end{figure*}

In \fig{fig:refPrincipal}, we show, for the ideal and the non-ideal cases, the evolution of density, temperature, velocity in the $x$-direction, total velocity, the $y$-component of the magnetic field, total magnetic field, Mach number, mass flux, ambipolar diffusion resistivity coefficient, and electron fraction as a function of the $x$-coordinate at three different times, for our reference case (see \tab{tab:params}). The same panels include a comparison with the ideal MHD case (dashed lines).

At early times, the density shows a peak with an initial maximum of about $3\times10^{-19}$~g~cm$^{-3}$ ($9\times10^{4}$~cm$^{-3}$) in the center, slightly increasing at later times. {The peak value of the $y$-component and the total magnetic field is higher than 100~$\mu$G}, triggered by the increase in density. In the outer regions, at 0.1$t_{\rm ff}$ the density is about $2\times10^{-21}$~g~cm$^{-3}$ ($6\times10^{3}$~cm$^{-3}$). The initial turbulence affects the first stages of the simulation generating fluctuations in the magnetic field, velocity, and temperature, the latter showing values of 20~K. {The density and velocity profiles both indicate that a shock front is generated at around $0.05-0.1$ pc from the filament center, after the initial stages}.

From the velocity field, we calculated the Mach number as $\mathcal{M} = \varv/c_{\rm s}$, where $\varv = (\varv_x^2 + \varv_y^2 + \varv_z^2)^{1/2}$ is the total velocity magnitude, and $c_{\rm s}$ the {current} sound speed of each cell. Similarly to the velocity field, the Mach number shows strong fluctuations at the first stages of evolution, going from values around 5 (highly supersonic) in the outer regions to $\mathcal{M}<1$ (subsonic) in the center. The same behavior is seen at later times.
From the density and the velocity field, we estimate the mass flux, reflecting the accretion flow in the $x$-direction:
\begin{equation}
        \Dot{\Sigma} = \rho \varv_x\,.
\end{equation}
At early times, the mass flux shows peaks of ${\sim10^3}$~M$_{\odot}$~Myr$^{-1}$~pc$^{-2}$ as a result of the density increase, which remains high also in the subsequent time-steps with a slight decrease.

The gravitational acceleration due to the mass accumulated in the filament leads to peak velocities up to 2 km~s$^{-1}$ moving inward, consistent with the expected free-fall  due to the available mass. The $y$- and $z$-components of the velocity (not shown), exhibit a similar behavior to the $x$-component, but reach peak values of $1$~km~s$^{-1}$ and $0.5$~km~s$^{-1}$, respectively, with the latter close to the edges of the simulation box. This relatively small difference in magnitude is caused by the initial statistics of the turbulence, that has the same dispersion on the three spatial components, but zero average only on the $x$-component.

The thermal profile, after an initial isothermal state, shows three distinct regions: ({\it i}\/) a cold background gas at around 10~K, ({\it ii}\/) an efficiently heated region with a temperature that is higher than 100~K produced by gas shocking, and ({\it iii}\/) the filament ridge where the high density leads to efficient cooling (i.e., cooling time shorter than the dynamical time). This brings the gas down to 10~K.

The $x$-component of the magnetic field remains constant in space and time by construction, while the $z$-component (not shown) is initially zero because of the alignment followed by fluctuations of a few $\mu$G due to the interaction with turbulence and velocity fluctuations. The $y$-component dominates the evolution of the total magnetic field, as seen in panels ({\it f}\/) and ({\it h}\/) of \fig{fig:refPrincipal}.

We note some differences from the ideal MHD case. In particular, over time, the density peak tends to be broader {in the ideal MHD case,} due to the increased magnetic pressure and stronger coupling between the magnetic field and the density.

\begin{figure*}
    \centering
    \includegraphics[scale=0.49]{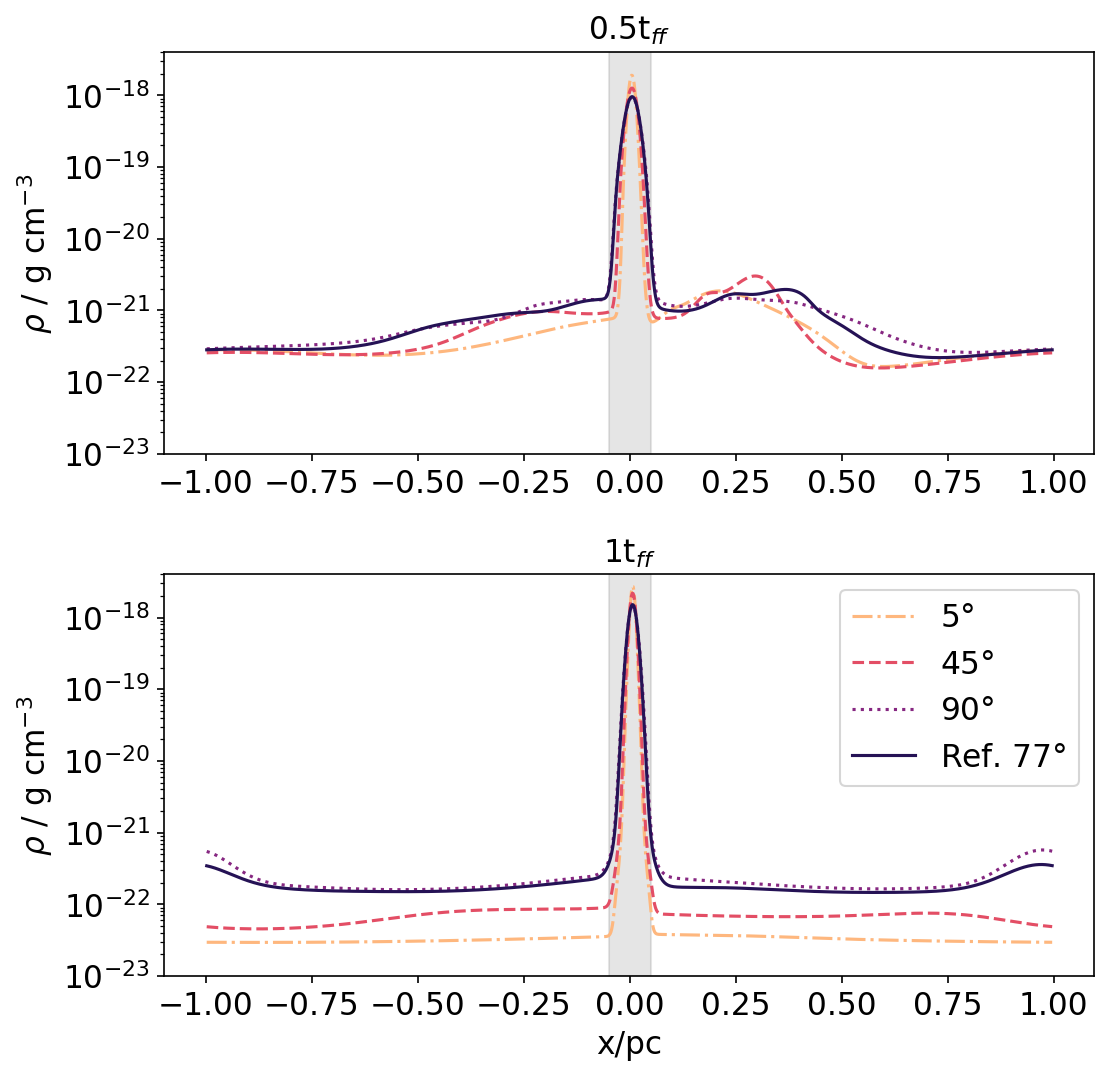}
    \includegraphics[scale=0.49]{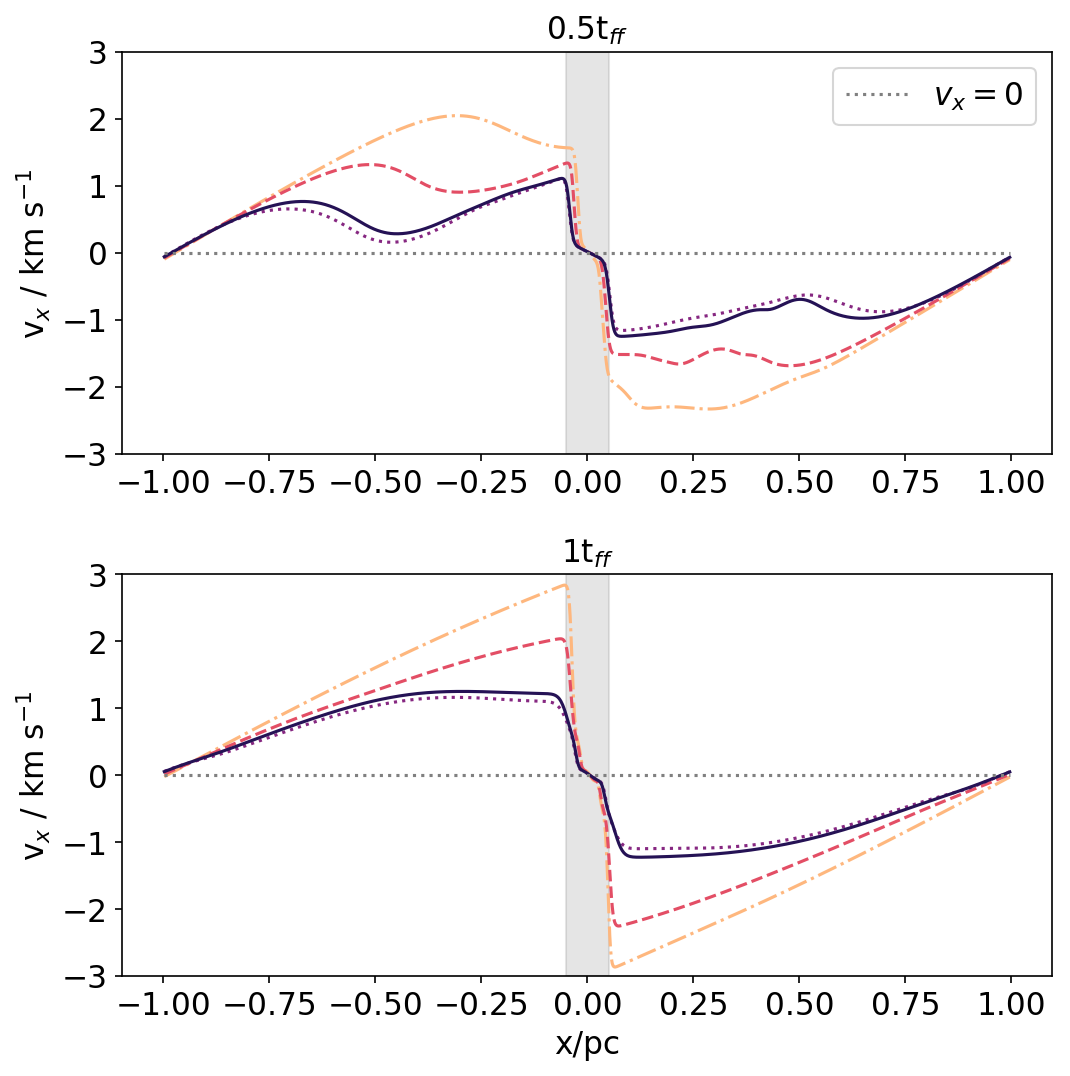}    
    \caption{Density and velocity along the $x$-axis for different initial angles. The solid line corresponds to the reference case. At 0.5$t_{\rm ff}$ in the top panels and  at 1.0$t_{\rm ff}$ in the bottom panels. The gray area in the middle corresponds to 0.1 pc to guide the reader to identify the typical width of the filaments.}
    \label{fig:3.2.rho and vx}
\end{figure*}

During the evolution, the density profile further steepens. At low densities, namely,~$\rho \lesssim 5\times10^{-20}$~g~cm$^{-3}$, the electron fraction (panel k) remains above $10^{-6}$, but as the density approaches a maximum value of about $10^{-18}$~g~cm$^{-3}$, the electron fraction drops to $\sim 10^{-10}$ due to efficient recombination. The decrease observed in the ambipolar diffusion at high densities ($n_{\rm H} \sim 10^5$~cm$^{-3}$) is determined by the electrons and X$^+$, which are the dominant charge carriers, while grains take over at $n_{\rm H} > 10^8$~cm$^{-3}$ \citep{Marchand2016}. The negatively charged grain abundance (not shown) is strictly linked to the electrons due to recombination; thus, the evolution of the two species is highly correlated. When the electron abundance drops off, the abundance of neutral and positively charged grains increases at high densities. Ambipolar diffusion keeps decreasing until the ion abundance increases enough for the diffusion with respect to the neutrals to become significant.

\subsection{Magnetic field geometry and strength} \label{magnetic field}

\begin{figure}
    \centering
    \includegraphics[scale=0.49]{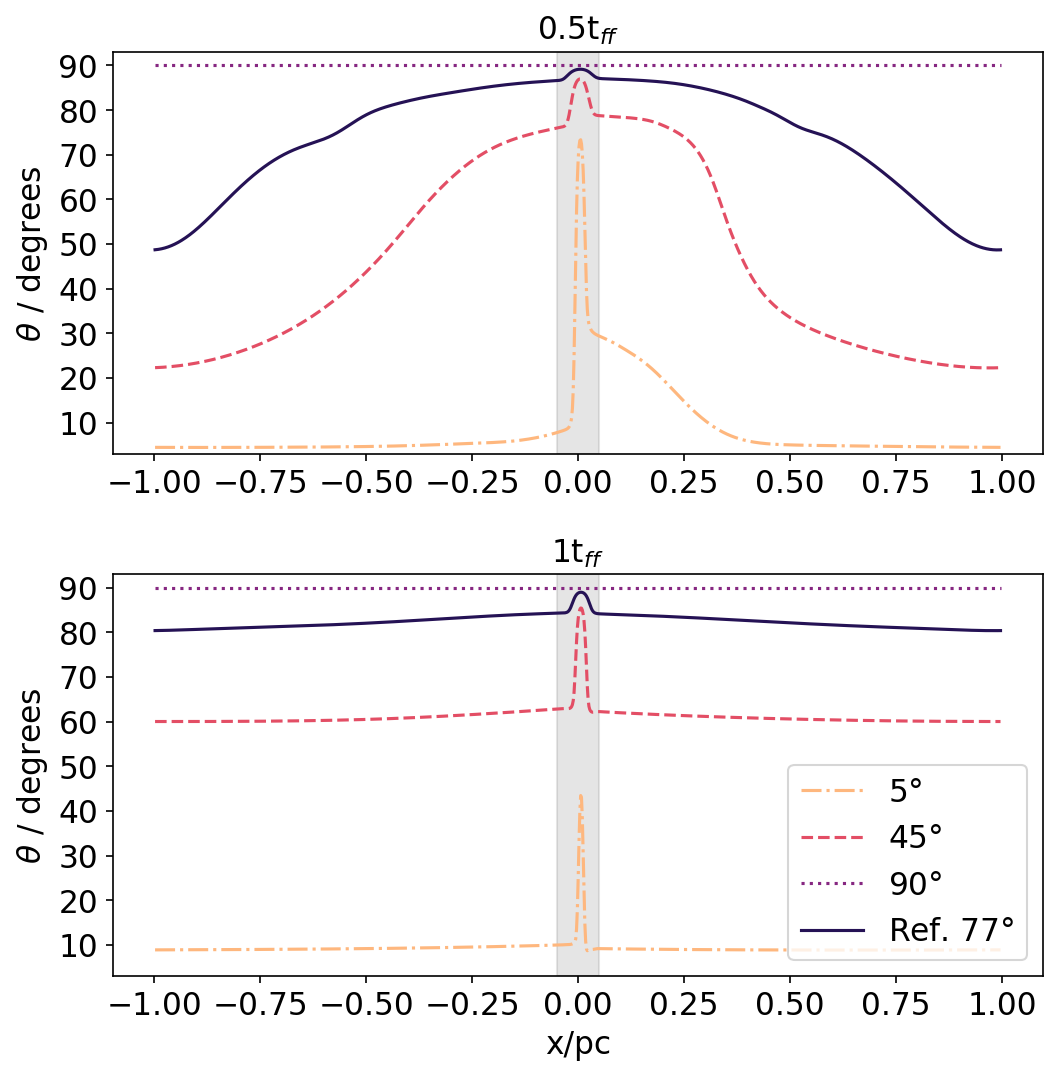}
    \caption{Inclination of the magnetic field and $y$-component for the different initial angles. The solid line corresponds to the reference case. We have $t=0.5 t_{\rm ff}$ in the top panels and  $t=1.0 t_{\rm ff}$ in the bottom panels. The gray area in the middle corresponds to 0.1 pc to guide the reader to identify the typical width of the filaments.}
    \label{fig:3.2.a}
\end{figure}

To study how the initial geometry of the magnetic field influences the evolution of the filament, we tested four different initial inclinations $\theta$ with respect to the $x$-axis, corresponding to $\theta=5\degree$ (almost parallel to the flow),  $\theta=45\degree$, $\theta=77\degree$ (reference case), and $\theta=90\degree$ (perpendicular to the flow){, while the other parameters are kept unchanged}. In \fig{fig:3.2.rho and vx} and \fig{fig:3.2.a}, the density, velocity and orientation angle of the magnetic field are shown at different times as a function of position. From now, we present the evolution of the model at two representative times: at half a free-fall time 0.5$t_{\rm ff}$, and after one free-fall time 1.0$t_{\rm ff}$. The central peak of the density and its width do not change very significantly for the different angles; However, when the magnetic field is nearly parallel to the gas flow, we would expect that the gas flow is strongly impeded by the magnetic field. As a result, the density in the outer parts becomes even lower when $\theta$ is small, without significantly changing  the density peak in the center, as the extra mass that is added to the center is very small. Similarly, the peak velocity increases as $\theta$ becomes aligned to the flow.

In general, the orientation of the magnetic field evolves both as a function of position and as a function of time. In particular, in the outer parts of the filament, the orientation angle decreases as a function of time and the magnetic field tends to align with the flow of the gas. In the filament itself, on the other hand, the velocities are reduced due to increased thermal pressure and the perpendicular component of the magnetic field is compressed and amplified as more gas is accreted onto the filament. As a result, the orientation becomes closer to perpendicular within the dense component of the gas. 

\begin{figure*}
    \centering
    \includegraphics[scale=0.49]{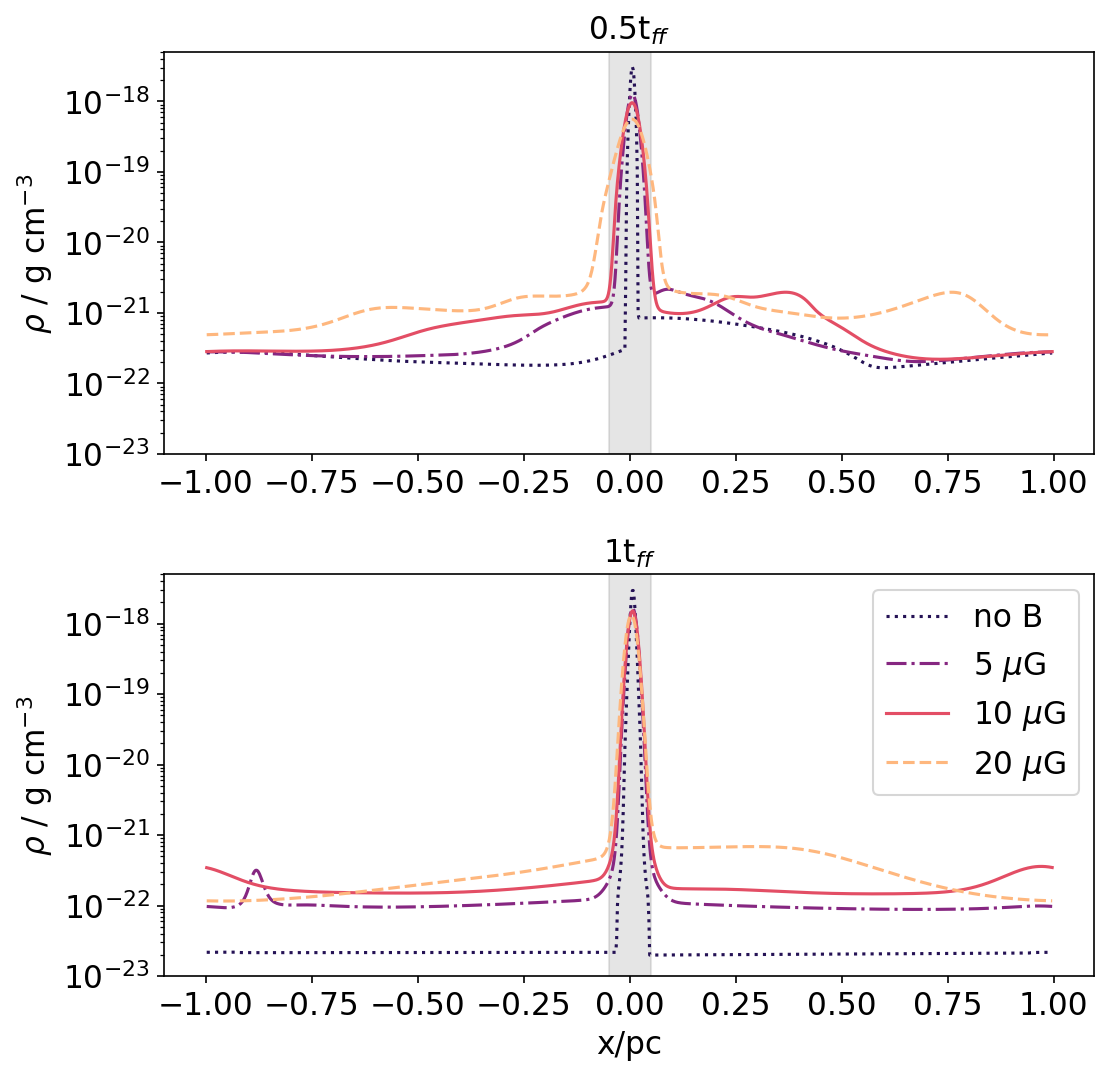}
    \includegraphics[scale=0.49]{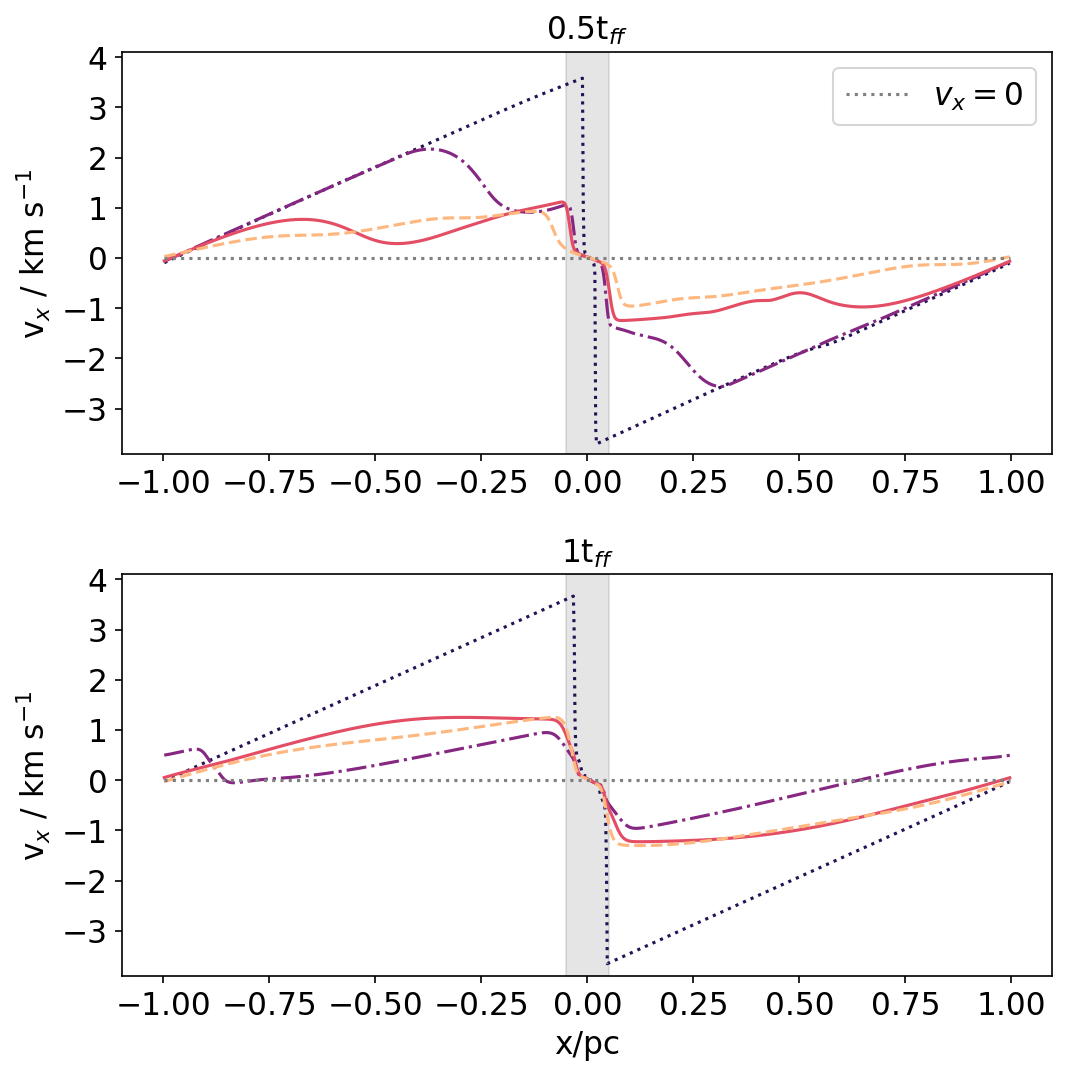}
    \caption{Density and velocity along the $x$-axis for the different initial magnitudes of the magnetic field. The solid line correspond to the reference case. We have $t=0.5 t_{\rm ff}$ in the top panels and  $t=1.0 t_{\rm ff}$ in the bottom panels. The gray area in the middle corresponds to 0.1 pc to guide the reader to identify the typical width of the filaments.}
    \label{fig:3.2-1.rho and vx}
\end{figure*}

It is important to note that because of the assumed symmetry, the simplified setup here may not necessarily hold in a realistic setting. Nevertheless, what we can see clearly from \fig{fig:3.2.a} is the change in the angle of the magnetic field between the outer and the inner parts of the filament. In a realistic setting, such a change of the angle may for instance correspond to a ``U''-shaped magnetic field, as suggested by \citet{Gomez2018}, or potentially other types of configurations that imply a change of the angle as a function of scale.

To study the effect of the magnetic field strength, we determined how the structure of the filament depends on four different initial field strengths, corresponding to no field (numerically $\sim10^{-25}$~$\mu$G), 5~$\mu$G (weak field), 10~$\mu$G (reference case), and 20~$\mu$G (strong field){, while the other parameters were kept unchanged}. {We calculated the normalized mass-to-flux ratio $(M/\Phi_B)/(M/\Phi_B)_{\rm crit}$ at the initial condition to show the magnetic supercritical nature of our simulations: $\sim$141 for 5~$\mu$G, $\sim$62 for 10 ~$\mu$G, and $\sim$31 for 20~$\mu$G}. In \fig{fig:3.2-1.rho and vx}, the density and the $x$-velocity are shown as a function of position. The magnetic field strength is one of the parameters that mostly affect the evolution of the filament, as highlighted after $0.5t_{\rm ff}$: as the magnetic field strength increases, the density peak of the filament appears smaller, while the filament appears wider. Because of the limited mass available, this also results in a background density that decreases (on average) for stronger magnetic fields (we note that because of the initial turbulent velocity, the local density in some cells does not necessarily reflect this trend). A simple explanation for this behavior is the increasing magnetic pressure, which counteracts gravity, slowing down accretion onto the filament, as can be also seen in the velocity profile, where the peak velocity is higher for weaker magnetic fields. This suppression results in higher densities in the background region, which also correspond to a higher average value of the magnetic field (despite the non-ideal effects) and lower inflow velocities (which arise from the combined effect of a smaller mass concentration producing a weaker gravitational pull and a larger pressure gradient counteracting it). At later times, these differences are almost washed out, since gravity dominates the system evolution, finally leading to an efficient accretion of material onto the central overdensity, as shown in the bottom panels. Nonetheless, mild differences in the density and velocity profiles remain in the underdense region outside the filament.

The density profile shows a similar trend at the earlier stages, whereas the velocity exhibits a peculiar evolution. In particular, the weakest magnetized case ($5~\mu$G) now has the lowest velocities. This result is due to the inversion from a magnetically dominated pressure to a thermally-dominated regime, where the previously higher inflow velocities produced a much stronger shock heating of the gas  near the filament, which then developed into a strong pressure gradient suppressing the inflow. In any case, these differences remain small, especially considering the corresponding gas densities.

\subsection{Cosmic ray ionization rate} \label{CRIR}
 
 \begin{figure*}
    \centering
    \includegraphics[scale=0.49]{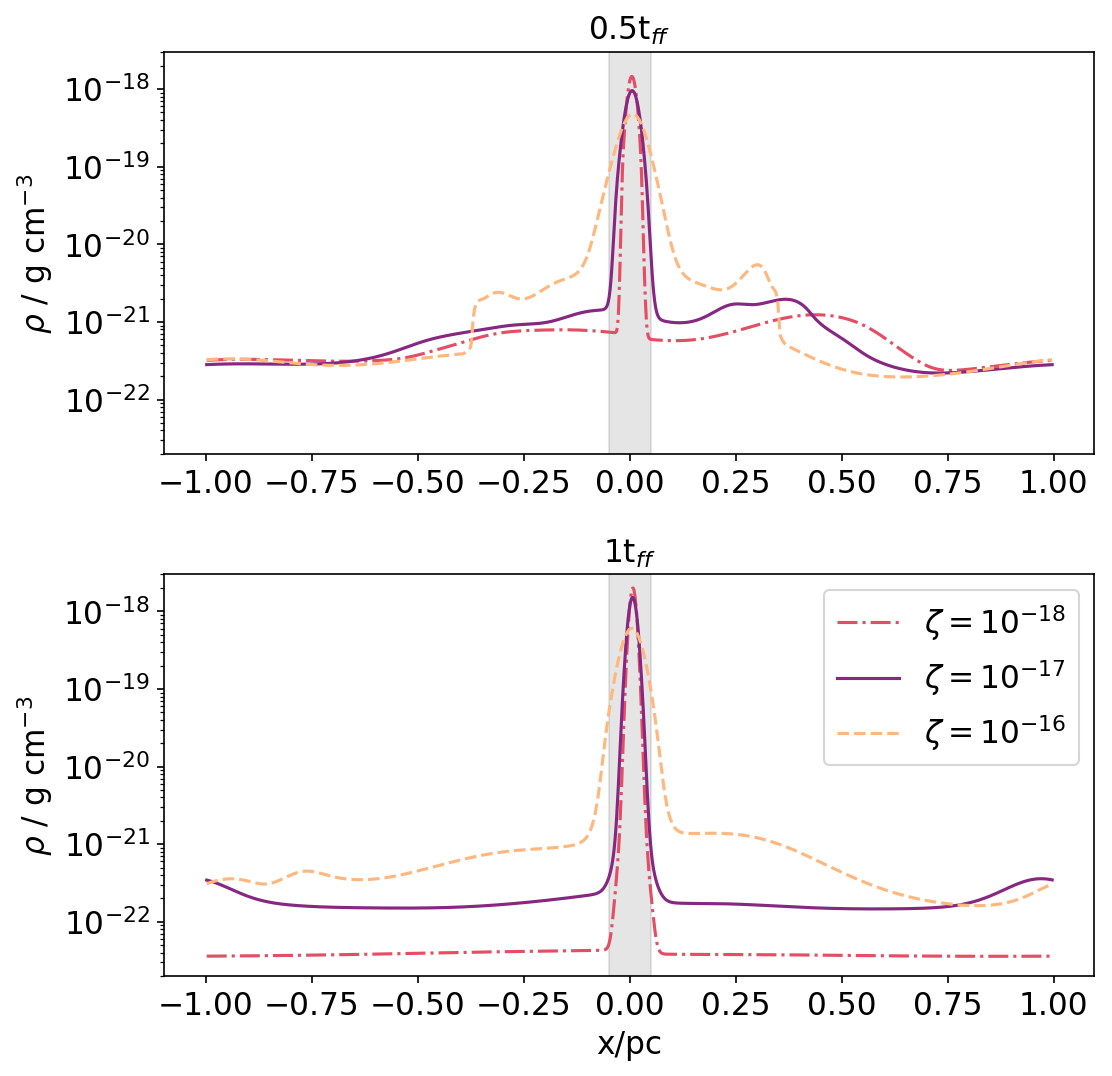}
    \includegraphics[scale=0.49]{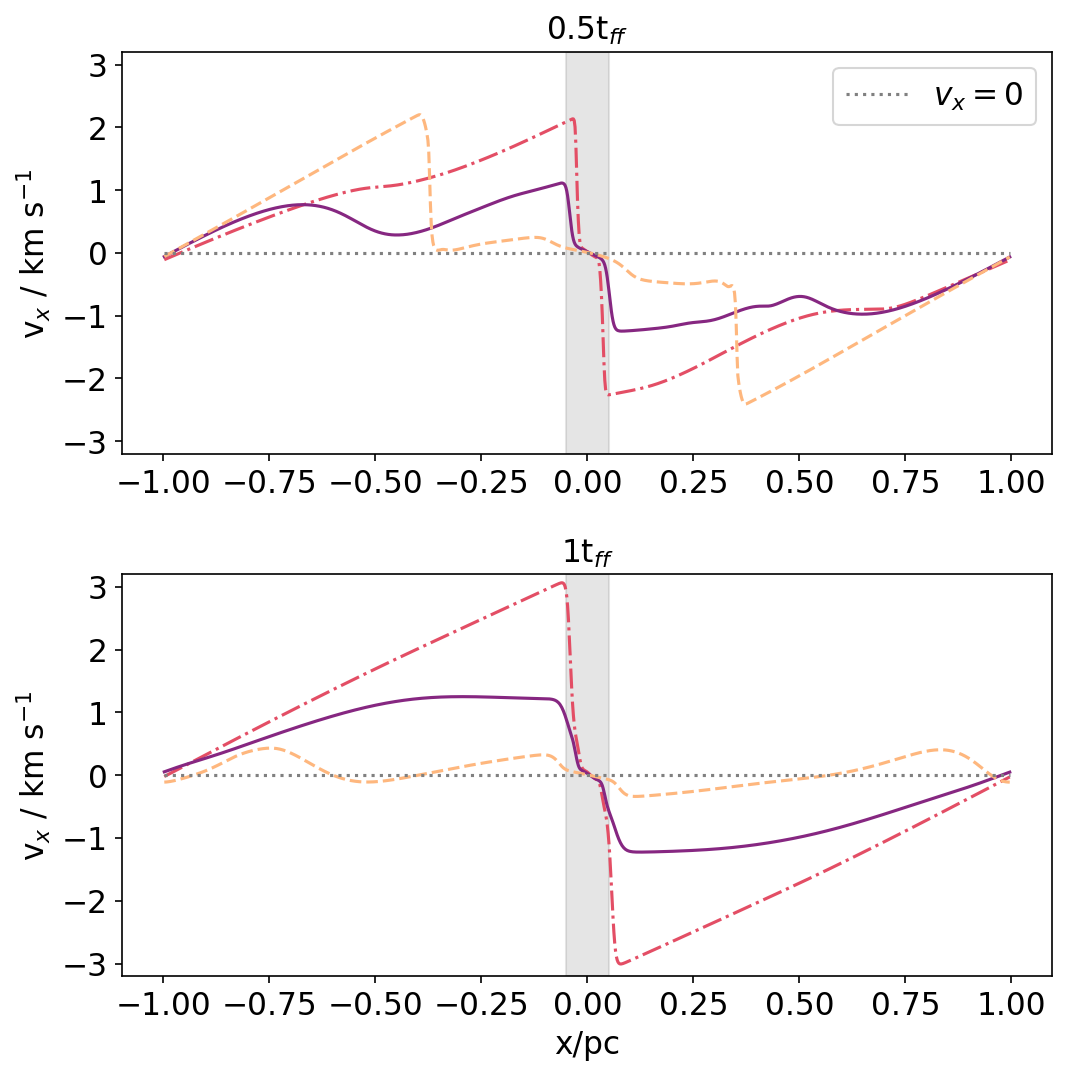}
    \caption{Density and velocity on the $x$-axis for different cosmic ray ionization rates. The black line correspond to the reference case. We have: $t=0.5 t_{\rm ff}$ in the top panels and  $t=1.0 t_{\rm ff}$ in the bottom panels. The gray area in the middle corresponds to 0.1 pc to guide the reader to identify the typical width of the filaments.}
    \label{fig:3.3.rho and vx}
\end{figure*}

\begin{figure*}
    \centering
    \includegraphics[scale=0.49]{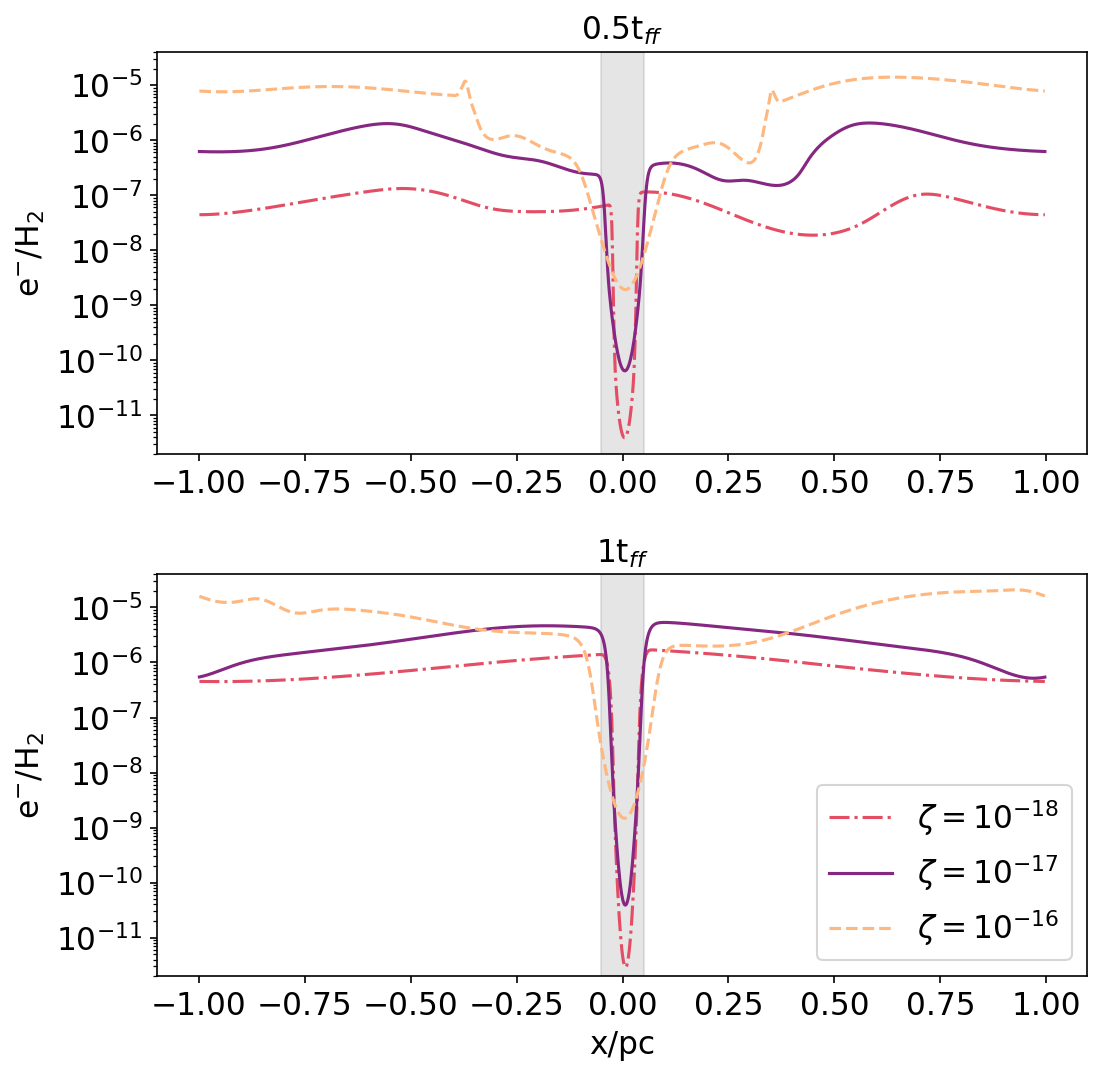}
    \includegraphics[scale=0.49]{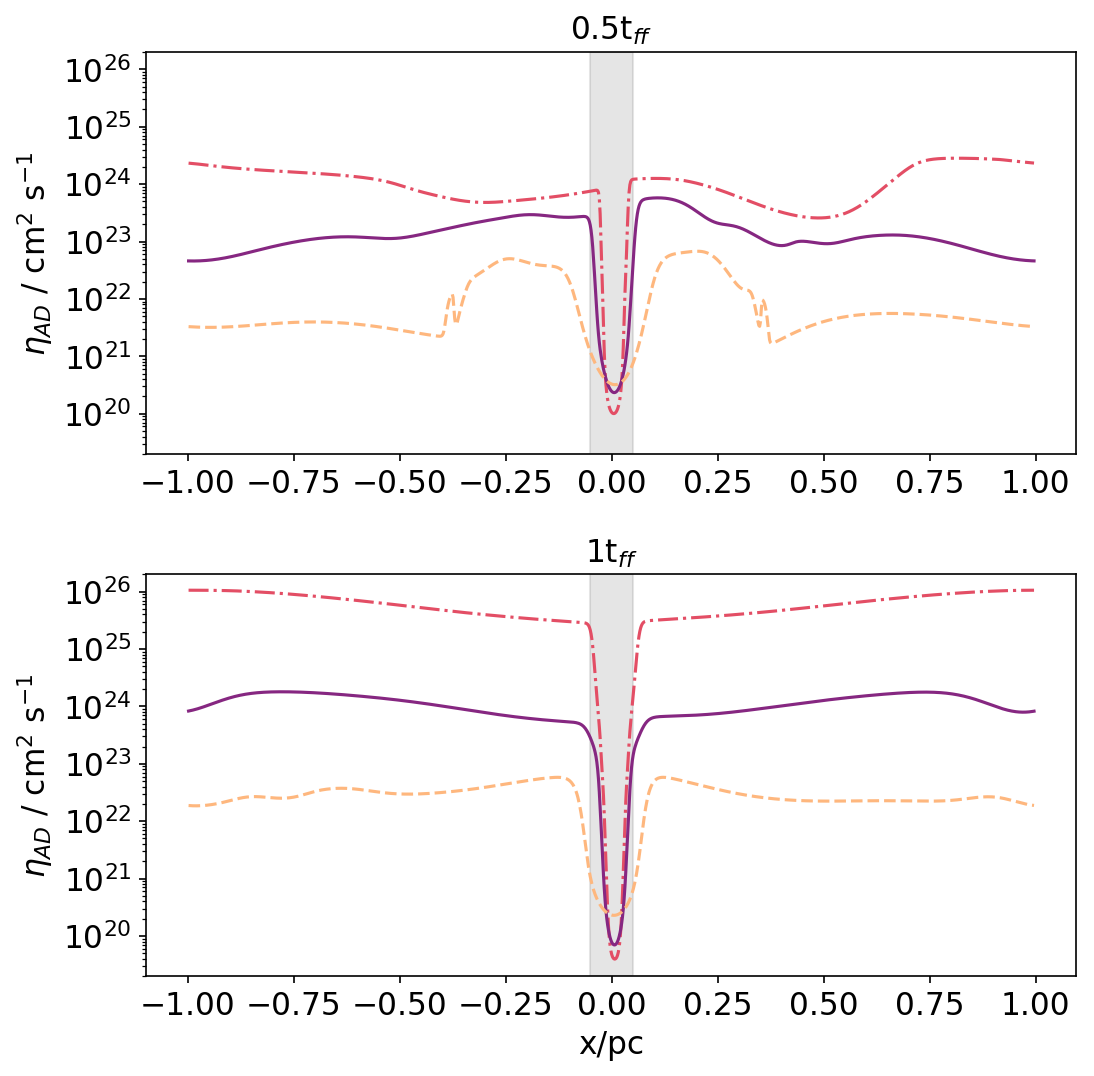}
   \caption{Ionization fraction and ambipolar diffusion resistivity coefficient for the different cosmic ray ionization rates. The black line correspond to the reference case. We have $t=0.5 t_{\rm ff}$ in the top panels and t=1.0 $t_{\rm ff}$ in the bottom panels. The gray area in the middle corresponds to 0.1 pc to guide the reader to identify the typical width of the filaments.}
    \label{fig:3.3.fion and eta}
\end{figure*}

To test the effect of the cosmic ray ionization rate (CRIR) on the evolution of the filament, we performed three simulations with $\zeta_{\rm cr} = 10^{-18}$~s$^{-1}$ (low), $10^{-17}$~s$^{-1}$ (reference), and $10^{-16}$~s$^{-1}$ (high), respectively{, while the other parameters were kept unchanged}. The density, velocity, electron abundance, and ambipolar diffusion resistivity coefficient are shown in \fig{fig:3.3.rho and vx} and \fig{fig:3.3.fion and eta} for the different cases and at two different times. Higher cosmic-ray ionization rates produce an increase in the ionization fraction in the low-density accretion region (see \fig{fig:3.3.fion and eta}, which corresponds to a stronger coupling between the gas and the magnetic field). As shown in \fig{fig:3.3.fion and eta}, the lower electron density (left panels) corresponds to a less-ideal MHD, namely,~a higher ambipolar diffusion coefficient (right panels), determining a direct relation between CRIR and non-ideal behavior.
Analogously, the density profile shows a broader distribution caused by a slow accretion toward the filament center, as shown in the right top panels of \fig{fig:3.3.rho and vx}. This behavior is also confirmed by the velocity profile at 0.5$t_{\rm ff}$, which shows in the high CRIR case a shift of the $\varv_x$ peak toward higher radii, determined by the relatively higher magnetic pressure support. This peak is no longer present at 1.0$t_{\rm ff}$, where the magnetic pressure halts the gravitational collapse, producing the slowest infall in the high CRIR case.

\subsection{Other parameters} \label{other parameters}

To complete our analysis, we studied the effects of additional parameters on the evolution of the filament. We have explored {independently} different initial Mach numbers, initial central density, the random seeds for the turbulence, and the exponent of the magnetic field-density relation, as in \eq{crutcher}, while the other parameters were kept unchanged.

By changing the initial turbulence seed, we did not observe relevant differences in the final evolution of the filament, suggesting that in our case the global features are dominating over the local turbulence, which dissipates over time. In fact, the turbulence-induced fluctuations (e.g.,~in the density profile) are more prominent at earlier stages.

For the different density regimes, we reduced $\rho_{\rm ridge}$ by a factor of 10 and 2, with the free-fall time increasing to $\sim$4.3~Myr and $\sim$1.9~Myr, respectively. As expected, at lower density it becomes harder for the filament to accrete material and there is no real evolution over time, with the density keeping around the ambient values. Similarly, reducing the density by a factor of 2 produces broader density profiles and a slower evolution. By changing the exponent of the magnetic field -- density relation from $k=0.5$ to $k=0.4$, \eqn{crutcher}, the filament reaches higher temperatures, velocity, and central density, but almost the same density peak with respect to the reference, since the lower magnetic pressure cannot slow down the accretion efficiently. Conversely, for $k=0.6$ (cf.~0.65 in \citealt{Crutcher2010}) the magnetic field is higher in both the ambient and the filament and it becomes more perpendicular to the flow compared to the previous cases, resulting in a slower accretion.
Although these parameters affect the global evolution, their effect is less pronounced than the effect of CRIR and of the initial magnetic field geometry.

\section{Discussion}

\begin{table*}
    \centering
    \begin{tabular}{ccccccccccc}
        \hline \hline
        & Reference & Ideal & $\theta=5$\degree & $\theta=45$\degree & $\theta=90$\degree & no-field & 5\,$\mu$G & 20\,$\mu$G & 10$^{-18}$\,s$^{-1}$ & 10$^{-16}$\,s$^{-1}$\\
        \hline
        \textlangle$\dot{\Sigma}$\textrangle / \textlangle$\dot{\Sigma}_{\rm ref}$\textrangle & 1 & 0.410 & 1.372 & 1.366 & 0.960 & 1.684 & 1.663 & 0.919 & 1.317 & 0.706\\
        FWHM/FWHM$_{\rm ref}$ & 1 & 7.214 & 0.575 & 0.698 & 1.046 & 0.534 & 0.922 & 1.159 & 0.755 & 2.447 \\
        \textlangle$\dot{\Sigma}$\textrangle & 206.15 & 83.828 & 282.87 & 281.55 & 197.92 & 347.08 & 342.802 & 189.48 & 271.58 & 145.58\\
        FWHM & 0.023 & 0.164 & 0.013 & 0.016 & 0.024 & 0.012 & 0.022 & 0.027 & 0.018 & 0.057 \\
        \hline
    \end{tabular}
    \caption{Time-averaged mass flux ratio, FWHM ratio, {time-averaged mass flux (M$_\odot$~Myr$^{-1}$~pc$^{-2}$) and FWHM (pc)} at t=1.0$t_{\rm ff}$, for an ideal case, the different inclination angles, magnetic field strengths, and cosmic ray ionization rates. Reference case have $\theta=77$\degree, $B_0=10\,\mu$G, and $\zeta=10^{-17}$\,s$^{-1}$. We note that $\theta=90\degree$ indicates that the magnetic field is perpendicular to the flow, and no-field corresponds to $B_0=0$\,$\mu$G.}
    \label{tab:accretion}
\end{table*}
In order to  assess the impact of the different parameters during the evolution of the gas, we should define one or more metrics that describe quantitatively the shape and the global morphological characteristics of the filament. The formal definition of its shape and width has been discussed by several authors, as we report in Appendix~\ref{apendix A}. Regarding the aims of our work, we found that our simulation results are better interpretable by defining two global metrics: (\textit{i}) the average accretion mass flux, $\langle\dot\Sigma\rangle$, and (\textit{ii}) the full-width-at-half-maximum, FWHM, obtained by fitting the density profile with a Gaussian function (see e.g.,~\citealt{Arzoumanian2011}) including the background up to $0.2$~pc {for a more suitable fitting of the curve}. The former is averaged over time, while the second is calculated at 1.0$t_{\rm ff}$, {that is, at the end of the simulation}. A summary of these two metrics {and} {their} relative ratio with respect to the reference case is reported in \tab{tab:accretion}.

As both metrics are controlled by the gas pressure support (either thermal or magnetic), the accretion mass flux and the FHWM appear to be anticorrelated: the FWHM increases for higher pressure, while the accretion rate tends to be reduced. This can be explained by observing the ideal MHD case, where the contribution of the magnetic pressure to the total pressure, reduces both the accretion and the capability of the filament to reach narrower configurations, resulting in larger FWHMs. In fact, in this case, the FHWM is 7~times larger than the reference case.  For the same reason, the CRIR cases present an increased FWHM in the high-CRIR model, and vice versa for the low-CRIR. The magnetic pressure plays a crucial role also when changing the initial magnetic field; reducing the field strength (no-$B$ and 5\,$\mu$G) narrows the density profile, while increasing the magnetic strength to $20$\,$\mu$G causes a relatively broader FWHM.

By changing the orientation of the initial magnetic field ($\theta$), we were able to observe a narrower FWHM when $B_y$ tends to be parallel to the flow (i.e.,~$\theta$=5\degree, almost perpendicular to the filament ridge). This decrease is driven by the magnitude of $B_y$, which is the only component that (by construction) varies spatially\footnote{$B_z$ is also variable, but it is less relevant in this context, being always smaller than $B_y$. Moreover,  $B_z$ is expected to remain zero except for the fluctuations induced by turbulence, that produce a zero average of $B_z$.}. When $\theta$ is small, $B_x$ is the dominant component of the magnetic field, namely, it is the component that (having no spatial gradient) plays a minor role in the MHD equations. In fact, in the right panels of \fig{fig:3.2.rho and vx}, $v_x$ is larger for the smallest $\theta$, being dominated by self-gravity and resulting in higher accretion and a narrower FWHM.

\begin{figure}
    \centering
    \includegraphics[scale=0.45]{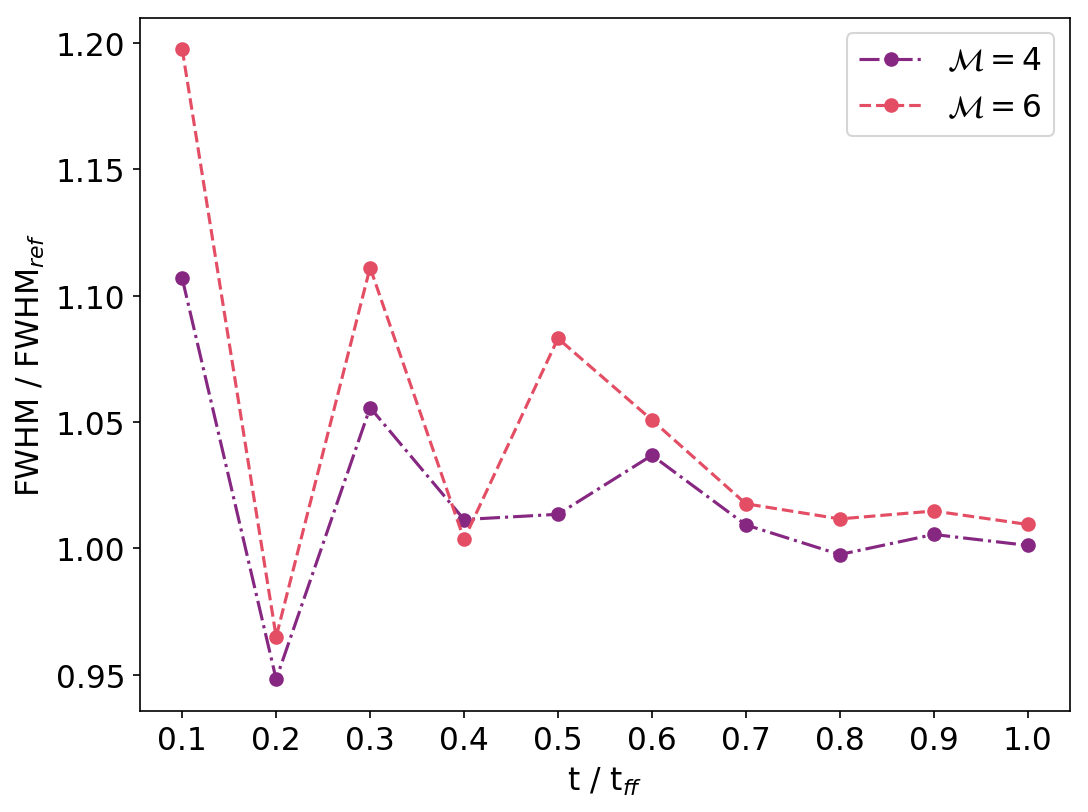}
    \caption{Time evolution of the FWHM ratio of $\mathcal{M}=4$ (dot-dashed line) and $\mathcal{M}=6$ (dashed line) with $\mathcal{M}=2$ (reference).}
    \label{fig:width M perp}
\end{figure}

To assess the impact of turbulence on the filament properties, we show in \fig{fig:width M perp}, the time evolution of the FWHM for different values of the Mach number for the reference case. High Mach-number cases tend to behave similarly to $\mathcal{M}=2,$ except at earlier times when turbulence is dominant. 

\section{Limitations of the model}

As in any numerical model, some approximations and assumptions need to be introduced in order to make the problem computationally tractable when including detailed microphysics and when a parameter study is planned. The main limitation in our approach is the 1D approximation employed to model the filament along the $x$ coordinate. Despite the code having three components for the vector variables, such as velocity and magnetic field, the $y$ and $z$ coordinates have periodic boundary conditions that limit the exploration of the spatial variability. Additionally, in a 1D code the solenoidal condition imposes by construction that the $x$ component of the magnetic field is constant in time and space. Another set of periodic boundary conditions is imposed at the boundaries of the simulation domain to avoid infinitely growing accretion onto the filament (outflow boundaries), or nonphysical configurations (Dirichlet boundary conditions, i.e., zero derivatives). Ideally, this could be avoided by using a much larger simulation box, where the boundaries remain unaffected during the simulation due to their distance from the central filament ridge. However, such a set-up requires much more computational resources or an adaptive mesh, which is beyond the aims of the present work.

Although the microphysics is limited by the reduced set of chemical reactions, it is still nevertheless capable of capturing the main features of larger chemical networks, especially the influence on the non-ideal behavior driven by ambipolar diffusion, as discussed in \citet{Grassi2019}. In addition, since the chemistry has a noticeable influence on the cooling functions, the current network might reduce the capability of our model to determine time-dependent effects of the thermal evolution. Finally, since our reduced chemical network cannot be used to determine the ratio between the atomic and molecular species accurately, we use a constant molecular weight and adiabatic index that might somewhat affect the results.
Additional limitations and method details are discussed in \citet{Grassi2019}, where the main features of the code are also presented.

\section{Summary and conclusions}
In this 1D study, we modeled an accreting self-gravitating turbulent filament using \textsc{lemongrab}, a non-ideal MHD code that includes chemistry and microphysical processes. We explored how the main parameters, namely, the~configuration and strengths of the magnetic field and the cosmic ray ionization rate, as well as other parameters such as the Mach number, initial turbulence level, initial ridge density, and the exponent of the magnetic field-density relation affect the evolution and accretion of the filament. Our main results can be summarized as follows:
\begin{enumerate}
    \item Including non-ideal MHD is crucial  with regard to the filament evolution. The ideal case produces a wider filament, with an FWHM that is approximately seven times greater than the non-ideal model, due to the increased magnetic pressure support. This suggests that non-ideal MHD is fundamental to understanding the accretion process.
    
    \item Independently of the initial configuration of the magnetic field, the magnetic field lines in the central part of the filament bend to reach a perpendicular configuration with respect to the $x$-axis (i.e., the accretion direction). This is consistent, for example, with ``U''-shaped model of \citet{Gomez2018}. 

    \item  Higher cosmic ray ionization rates produce higher ionization degrees that correspond to a more ideal gas. As the coupling with the magnetic field is stronger, the magnetic pressure halts the collapse, resulting in a broader filament.
    
    \item We did not find any strong change in the final FWHM value with the Mach number, pointing to a less pronounced role of the turbulence in regulating the evolution and the final properties of the filament compared to microphysics and the magnetic pressure. However, a one-dimensional model could be inadequate to address this specific question.
\end{enumerate}

We conclude that adding magnetic fields and non-ideal MHD effects significantly affects the evolution of a collapsing filament, its width, and its accretion rate, while other parameters play only a minor role. Special attention should be given to the cosmic ray ionization rate that strongly affect the coupling between the gas and the magnetic field. It is then fundamental for future works to include  a proper cosmic rays propagation scheme to accurately study their effect on the accretion rate and on the filament width.

\begin{acknowledgements}
We are grateful to the referee for the detailed report. NGV acknowledge support by the National Agency for Research and Development (ANID) / Scholarship Program / Mag\'ister Nacional/2020 - 22200413. DRGS thanks for funding via Fondecyt Regular (project code 1201280). SB and DS gratefully acknowledge support by the ANID BASAL projects ACE210002 and FB210003. AL acknowledges funding from MIUR under the grant PRIN 2017-MB8AEZ. TG acknowledges the financial support of the Max Planck Society.

\end{acknowledgements}

\bibliographystyle{aa}
\bibliography{references}

\begin{appendix}
\section{Boundaries of the filament} \label{apendix A}
\begin{figure}
    \centering
    \includegraphics[scale=0.45]{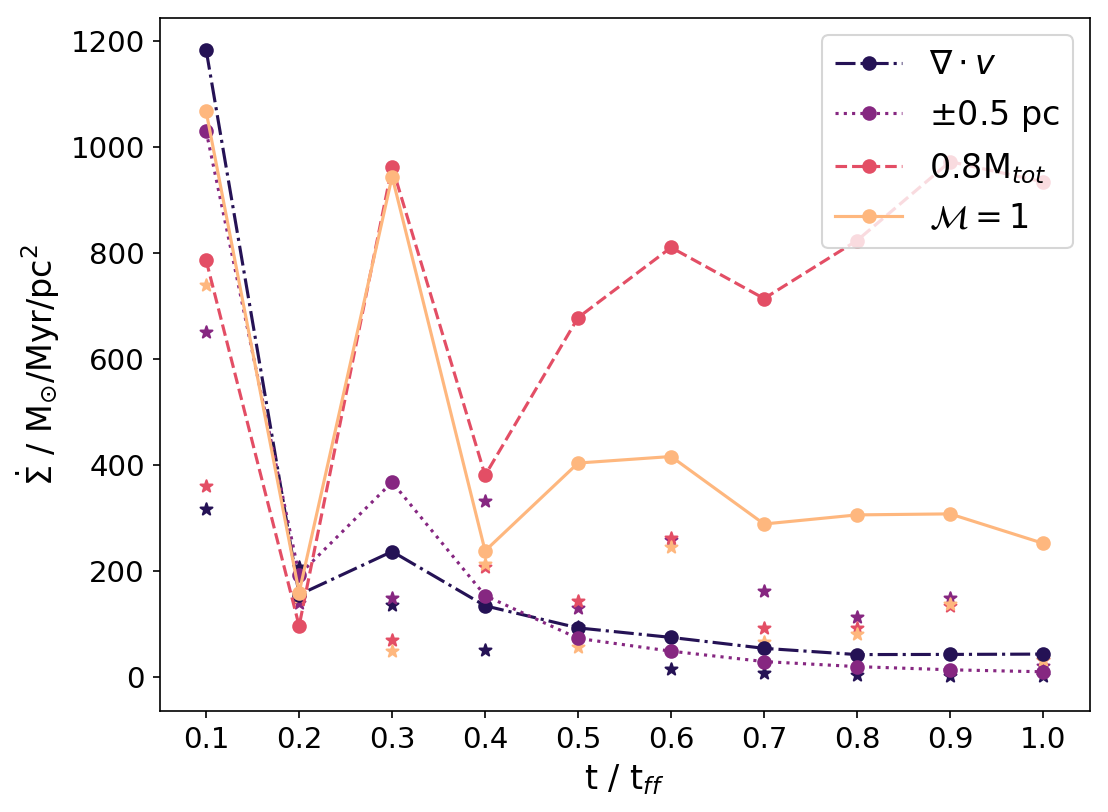}
    \caption{Time evolution of the mass flux according to the different methods adopted here. Circles correspond to non-ideal MHD, stars to ideal MHD.}
    \label{fig:4.1.mdot}
\end{figure}

As there is no unique definition of the boundaries of a filament, we calculated the mass flux for the reference case at different distances from the filament axis to make a comparison with the observational data. These positions are given by: ({\it a}\/) the minimum values of the velocity-divergence (or $x$-velocity gradient), as in \citet{Priestley2022a}; ({\it b}\/) $x=\pm 0.05$~pc from the axis of the filament (half of the filament width reported by \citealt{Arzoumanian2011}); ({\it c}\/) the distance where the enclosed mass is 80\% of the total mass in the computational box; and {(\it d\/)} where the flow becomes supersonic with a possible shock with Mach number $\mathcal{M}=1$. \fig{fig:4.1.mdot} shows the evolution of the mass fluxes in each time-step for each case, where dots correspond to non-ideal MHD and stars to ideal MHD. By looking at the ideal case (stars) we note that the difference is not very large after 0.3$t_{\rm ff}$. The total averaged mass fluxes in time for non-ideal MHD are 206.15~M$_{\odot}$~Myr$^{-1}$~pc$^{-2}$ for the criterion ({\it a}\/); 194.02~M$_{\odot}$~Myr$^{-1}$~pc$^{-2}$  for the criterion ({\it b}\/); 715.87~M$_{\odot}$~Myr$^{-1}$~pc$^{-2}$  for the criterion ({\it c}\/); and 438.38~M$_{\odot}$~Myr$^{-1}$~pc$^{-2}$ for $\mathcal{M}=1$  for the criterion ({\it d}\/).

Observational studies, such as in \citet{Palmeirim2013}, provided an estimate of the accretion rate in the B211 filament based on the observed mass per unit length and the $^{12}$CO (1--0) inflow velocity, finding a value of $\sim 27$--50~M$_{\odot}$~Myr$^{-1}$~pc$^{-1}$. With this accretion rate, it would take $\sim 1$--2~Myr to form the B211 filament, in reasonable agreement with the free-fall time of $\sim 1.3$~Myr of our model. Since the accretion rate is estimated in the observations at a distance of 0.4~pc from the filament axis, the method of the width is not appropriate for a comparison, because of the smaller radius used and because the width of the filament changes with time. The methods based on Mach number and $0.8 M_{\rm tot}$ are not well suited for a comparison, as they assume a position too close to the filament center. On the other hand, the method of the velocity-divergence (which is the chosen method) displays a greater similarity to the observations, where the inflow velocity is $\sim 0.6$--1.1~km~s$^{-1}$; whereas in our model, the velocity in the $x$-axis at the free-fall time $t=t_{\rm ff}$ is $\sim 1.2$~km~s$^{-1}$ at $\sim 0.1$~pc from the center of the filament. 
\end{appendix}
\end{document}